\def\year{2017}\relax
\documentclass[letterpaper]{article}
\usepackage{aaai17}
\usepackage{times}
\usepackage{helvet}
\usepackage{courier}
\usepackage[draft]{hyperref}
\usepackage{todonotes}
\usepackage{soul}
\usepackage{csquotes}
\usepackage{caption}
\usepackage{floatrow}
\begin{document}
\title{The Aftermath of Disbanding an Online Hateful Community}
\author{}
\author{Haji Mohammad Saleem \and Derek Ruths\\
School of Computer Science\\
McGill University\\
Montreal, Canada\\
haji.saleem@mail.mcgill.ca \and derek.ruths@mcgill.ca\\
}

\maketitle
\begin{abstract}
Harassing and hateful speech in online spaces has become a common problem for platform maintainers and their users.  The toxicity created by such content can discourage user participation and engagement. Therefore, it is crucial for and a common goal of platform managers to diminish hateful and harmful content. Over the last year, Reddit, a major online platform, enacted a policy of banning sub-communities (subreddits) that they deem harassing, with the goal of diminishing such activities.  We studied the effects of banning the largest hateful subreddit (r/fatpeoplehate or FPH) on the users and other subreddits that were associated with it. We found that, while a number of outcomes were possible --- in this case the subreddit ban led to a sustained reduced interaction of its members (FPH users) with the Reddit platform. We also found that the many counter-actions taken by FPH users were short-lived and promptly neutralized by both Reddit administrators and the admins of individual subreddits. Our findings show that forum-banning can be an effective means by which to diminish objectionable content.  Moreover, our detailed analysis of the post-banning behavior of FPH users highlights a number of the behavioral patterns that banning can create.
\end{abstract}

\section{Introduction}
Hate speech, generally defined as disparaging speech against minorities based on their race, ethnicity, sexual orientation, gender and religion \cite{mendel2012does}, has become endemic to online platforms with user-generated content\cite{bartlett2014anti}. Comprehensively, hate speech is an ``expression of hatred against a person based on their group identity" \cite{saleem2016web}.  Online hate speech can take several forms: for example, a microblog that verbally abuses a black actor based on her race\cite{woolf2016leslie}, comments under news articles\cite{hughey2013racist} or in-game chat rooms of major MMOGs (Massively Multiplayer Online Games)\cite{powers2003real}.

From the perspective of platform managers, hateful speech creates negative user experiences, which can impact the civility of their platform\cite{santana2014virtuous}\cite{newell2016user}. At the user level, the effects of online hateful speech can extend far beyond the virtual world, causing serious psychological distress\cite{boeckmann2002hate}\cite{leets1997words}. In extreme cases, hate speech can lead to physical harm through violence incited against the targets of the hate\cite{ushmm2010hate}. As a result, diminishing hateful speech in online social spaces has become a major objective of both platform maintainers and many of the users who use them.

To combat such objectionable content, platforms employ several strategies ranging from community-based flagging to platform-level banning \cite{grimmelmann2015virtues}\cite{chancellor2016thyghgapp}. Despite the widespread use of such techniques, little is known about the implications of these policies: in other words, even though platforms routinely deploy countermeasures against hate speech and their producers, very little is known about the direct and indirect effects of such actions - on the hateful content and on the communities affected by it.

In this study, we aim to understand one common practice among online forums: community banning.  We specifically consider this in the context of Reddit and its policy, enacted in 2015, of banning ``subreddits that allow their communities to use the subreddit as a platform to harass individuals when moderators don't take action''\cite{reddit2015removing}.

Reddit is a social media content aggregation and discussion website, which works on the model of multiple sub-communities called ``subreddits''. Subreddits cover a wide variety of topics, and users can subscribe to the subreddits that interest them or create their own subreddits to personalize their Reddit experience. Up until the summer of 2015, Reddit allowed users to create subreddits dedicated to hateful themes. r/fatpeoplehate (FPH), for example, was one such subreddit that focused on body shaming.

The FPH subreddit was highly popular, with approximately 150,000 subscribers at its peak (Figure 1). In 2015, Reddit introduced a new policy to ban harassing subreddits. \texttt{r/fatpeopelhate} was the most prominent subreddit banned under this policy. This study investigates the after-effects of banning this large self-identifying hateful subreddit on the users that participated in it (FPH users) and other subreddits that are related to it. Our analysis is restricted to FPH since it was the only example of a large-scale active subreddit disbanded by the policy update.

A priori banning a subreddit hardly guarantees that the kind of hateful content produced by the community will go away.  Several outcomes are possible.  One outcome involves the community simply creating a new subreddit under a different name.  Another outcome can be that hateful speakers post their content in other subreddits, making the hateful content much harder to expunge and possibly directly exposing target users to the hateful content.

Our work in this study produced strong evidence that all of these negative outcomes happened - but were very short-lived.  Ultimately, we found that the banning of the FPH subreddit reduced the engagement of active FPH contributors with the Reddit platform and dramatically decreased the volume of their overall comments. Not only did the FPH users comment less after the ban, a larger portion of the users stopped their commenting activity entirely. Furthermore, the initial response of the banned users (creating new subreddits and posting their hateful content to other platforms) was quickly neutralized (within 1-2 days) by the combined actions of Reddit administrators and the moderators of specific subreddits. The fact that FPH users were denied the opportunity to create a new community and were not welcomed, en mass, into existing communities, can explain their disengagement with the platform.

It is not possible to definitively conclude that banning of the subreddit reduced the presence of body-shaming related hateful content on Reddit. However, our overall findings strongly suggest that banning of an objectionable community significantly decreased the future participation of the active members of the community with the Reddit platform. Notably, achieving this end involves swift, independent, and decisive action on the part of both administrators and moderators.
\section{Background}

\subsection{Hate speech}
At the outset, establishing precisely what constitutes hate speech is an unsolved and important problem. In order to address often-contested definitions of hate speech and remove ambiguity, Saleem et al.\cite{saleem2016web} introduced the term ``hateful speech'' which they define as ``speech which contains an expression of hatred on the part of the speaker/author, against a person or people, based on their group identity''. This definition removes the ambiguity in the term ``hate'' itself, which might refer to the speaker/author's hatred, or his/her desire to make the targets of the speech feel hated, or their desire to make others hate the target(s), or the apparent capacity of their speech to increase hatred. Therefore, in this article we use the term ``hate speech'' in a general sense and the term ``hateful speech'' specifically as defined above. Furthermore, body-shaming speech adheres to the definition of hateful speech as it constitutes disparaging a group of people based on their appearance.

\subsection{The harmful effects of body-shaming speech}

Like other forms of hate speech, body shaming can have severe adverse effects on its victims. In adolescents, body-shaming was associated with lower body satisfaction, lower self-esteem, higher depressive symptoms, and thinking about and attempting suicide, even after controlling for actual body weight\cite{eisenberg2003associations}.  Furthermore, it was found that over-weight women significantly increase their calorie intake on exposure to weight-related stigma\cite{schvey2011impact}, causing stress, reducing self-control and ultimately resulting in further weight gain\cite{major2014ironic}\cite{tomiyama2014weight}. Therefore, body-shaming can lead to significant negative effect on the physical and mental well-being of plus-sized individuals.

\subsection{Banning of /r/FatPeopleHate}

\texttt{/r/FatPeopleHate} (FPH) was a subreddit created to mock and denigrate plus-sized people. The users would often post pictures of plus-sized individuals and ridicule them along with the usage of slurs, such as ``hamplanet'', ``landwhale'', ``beetus''. One of the rules for contributing to the subreddit included having ``absolutely no fat sympathy''. While the community admitted that it may seem that ``all we do is aimlessly bully and ridicule people", they maintained that their actions have a deeper meaning. They go on to state that obesity is a choice that indicates selfishness, lack of discipline and causes disproportionate burdens on health care systems.

\begin{figure*}
  \centering
  \includegraphics[width=0.7\columnwidth]{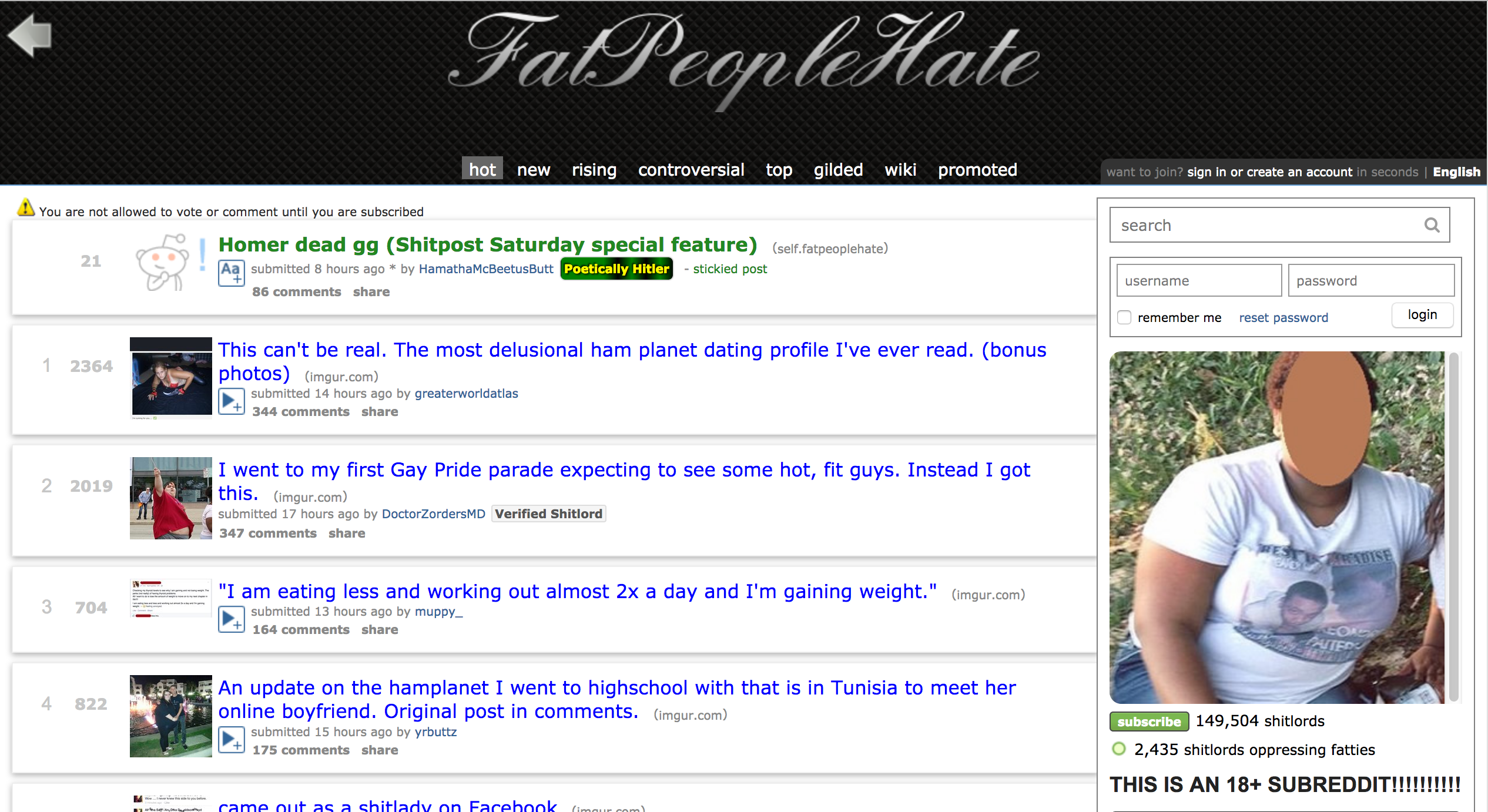}
  \caption{\textbf{A screen shot of \texttt{/r/fatpeoplehate} obtained through WayBack Machine dated June 7, 2015.} It shows how users would link images of plus-sized individuals accompanied with mocking titles. It also shows the number of subscribers at that moment: 149,504.}~\label{fig:fph_ss}
\end{figure*}

The subreddit grew to become very popular and had almost 150,000 subscribers at its peak. A screen shot of the subreddit is provided in Figure \ref{fig:fph_ss}, obtained through WayBack Machine\footnote{\url{https://web.archive.org/web/20150607141552/http://www.reddit.com/r/fatpeoplehate/}}. On June 10, 2015, Reddit admins announced that they were going to remove harassing subreddits \cite{reddit2015removing}. They clarified that ``we will ban subreddits that allow their communities to use the subreddit as a platform to harass individuals when moderators don't take action. We're banning behavior, not ideas.''. This announcement marked the implementation of new community policy that allowed Reddit to ban or quarantine objectionable subreddits. While some people supported the decision, others saw it as against free speech. 

In this paper, we present the reaction of the users affected by this ban and try to gauge the success of Reddit's community policy in controlling FPH-specific hateful speech on their platform. 

\section{Data}
Reddit user Jason Baumgartner, under the Reddit user name \texttt{Stuck\_In\_the\_Matrix}\footnote{\url{https://www.reddit.com/user/Stuck\_In\_the\_Matrix/}}, has made available large data dumps that contain a majority of the content (posts and comments) generated on Reddit\footnote{\url{http://files.pushshift.io/reddit/}}. The data dumps, collected using the Reddit API, are organized by month and year dating back to 2006. This resource is regularly updated with new monthly content. 
For our research, we focused on June 10, 2015 as the pivot point since it was on this day that Reddit announced the banning of harassing subreddits, the most prominent of which was \texttt{/r/fatpeoplehate}. Since we were interested in understanding the effects of banning such a large community, we studied the activity of users primarily associated with the FPH subreddit, over a four-month period, two months before and two months after it was banned. Thus, we restricted our analysis to the time period between the days of April 10, 2015 and August 10, 2015. An important note, the data dumps miss about 0.03\% of the public comments due to data crawling errors. Our analysis of macro-level trends should remain unaffected by the data-loss. 

\noindent{\bf FPH users.} For this time period, we collected all users that commented in \texttt{/r/fatpeoplehate}. This provided us with a list of 42,354 users. From this list we removed common bot accounts, such as \texttt{/u/AutoModerator, /u/autotldr, /u/Mentioned\_Videos}, etc. When users delete their accounts or the moderators remove comments, the authors of such comments would be displayed as \texttt{[deleted]}. We removed these from our list as well, since there is no way to link such comments to their original authors.  In addition, this initial sample presumably included some (small) population of users who do not condone the FPH community's views but still comment in the subreddit for the sake of discussion or argument. In order to remove such accounts, we only considered users who had FPH as their most commented subreddit. This left us with 13,916 users, whom we refer to as ``FPH users'' in this paper. For each FPH user, we collected all the comments associated with them in the period of interest and call it the ``FPH sample''.

\noindent{\bf Random users.} We sought to compare the activity of FPH users with an equal-sized systematic random sample of Reddit users. To this end, we curated a list of all the Reddit users that posted a comment during the period of interest. After removing the major bot accounts and FPH participants, we had a list of 4,677,759 users. To obtain an equally sized subset we perform systematic sampling. To do so, we first sort our list of users by the number of comments they made during this period. We selected an initial user randomly from the list and the remaining users at regular interval from the last one, to obtain a sample of 13,916 users (``random users''). Again, for each random user, we collected all the associated comments in the relevant time period and call it the ``random sample''. The details on the two samples are provided in Table \ref{tab:table1}.

\begin{table}
\centering
\begin{tabular}{l r r}
\small{Sample}      & \small{\# of users}   & \small{\# of comments}    \\
\hline
FPH                 & 13,916                & 878,276                   \\
Random              & 13,916                & 585,632                   \\
loseit              & -                     & 159,250                   \\
fatlogic            & -                     & 266,456                   \\
fatpeoplestories    & -                     & 45,484
\end{tabular}
\caption{\textbf{Details on each of our data sample.} Note that the three communities we investigate also differ in size, which can induce different behaviours.}~\label{tab:table1}
\end{table}

\noindent{\bf Subreddit-specific comments.} We were also interested in studying the effects of the ban on other communities, especially the ones with themes related to the banned community. Accordingly, we collected all the comments generated in:
\begin{description}
\item [\texttt{r/loseit}:] a subreddit that supports people who want to lose weight.
\item [\texttt{r/fatlogic}:] a subreddit that presents itself as against the ``fat way'' of thinking and not being a hate sub. They add ``\texttt{r/fatpeoplehate} is the place for that.''
\item[\texttt{r/fatpeoplestories}:] a subreddit to share the stories about fat people. They clarify that ``We are NOT FatPeopleHate''. 
\end{description}

Further details are provided in Table \ref{tab:table1}.

\section{Study Design}
At a high-level, we sought to characterize the effects of banning a prominent community on Reddit in two ways: 

\begin{enumerate}
    \item at the user level, we wish to study the effects on the active members of the banned community, and
    \item at the community level, we want to study if the ban created any counter effects on other communities, specifically those that operate in a similar sphere as the banned community.
\end{enumerate}

\noindent{\bf Banning and users.} We analyzed multiple aspects of FPH user population behavior, contrasting it with behavior exhibited by the random user group:

\begin{enumerate}
    \item the difference in overall commenting behaviour between the two samples over our period of analysis; 
    \item the difference in the commenting behaviour of the two user groups before and after the ban;
    \item the retention of users by the platform after the banning of the subreddit;
    \item the effect on bad user behaviour in the form of downvoted content;
    \item the change in user exploration patterns of new subreddits; and
    \item the search for a an alternative subreddit for the disbanded community.
\end{enumerate}

\noindent{\bf Banning and other communities.} We also studied the direct effects of the ban on other communities that are relevant to the topic of the banned subreddit. Specifically, we studied \texttt{r/loseit, r/fatlogic and r/fatpeoplestories}. For each of these subreddits, we studied the downvoted comment volume, the deleted comment volume, and the overall comment volume during our four-month period of interest.  As will be discussed later, these metrics were chosen as proxies for the presence of FPH-specific negative content in other subreddits before and after the ban.

\section{Results}

\begin{figure}[!b]
\centering
  \includegraphics[width=0.67\columnwidth]{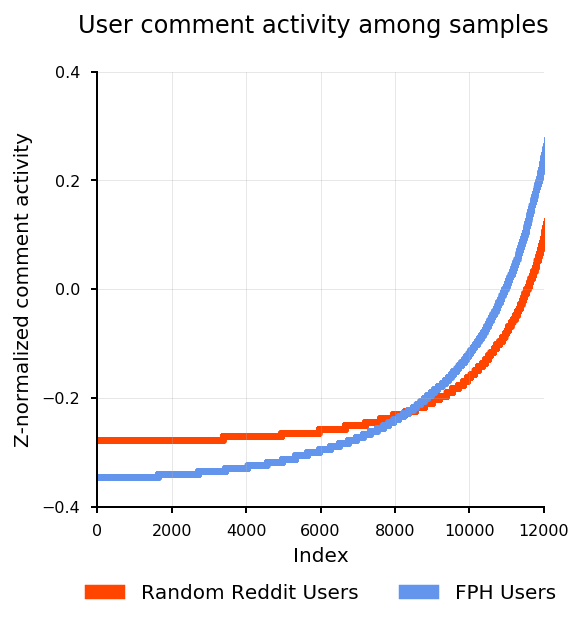}
  \caption{\textbf{Z-normalized comment activity of the two samples.} FPH and random users are ranked based on their total comment activity. Then the Z-normalized comment activity is plotted at each rank with FPH users indicated in blue and random users in orange. This figure contains the first 12,000 users for a clearer representation.}~\label{fig:totalcomms}
\end{figure}

\noindent{\bf The overall commenting behaviour of the FPH sample is significantly different from that of the random sample.} Figure \ref{fig:totalcomms} presents the difference in overall commenting behaviour between the two samples during the period of analysis. The two samples follow each other very closely, which is backed by the Pearson's correlation coefficient of 0.97. Comparing the two distributions by the Kolmogorov-Smirnov goodness-of-fit test returns a p-value of <0.0001. This indicates that the null hypothesis that the two independent samples are drawn from the same continuous distribution can be rejected, and that there is a significant difference in the overall behaviour of users in our two samples.

\begin{figure*}
\begin{center}
\begin{tabular}{c c c}
\includegraphics[width=0.32\textwidth]{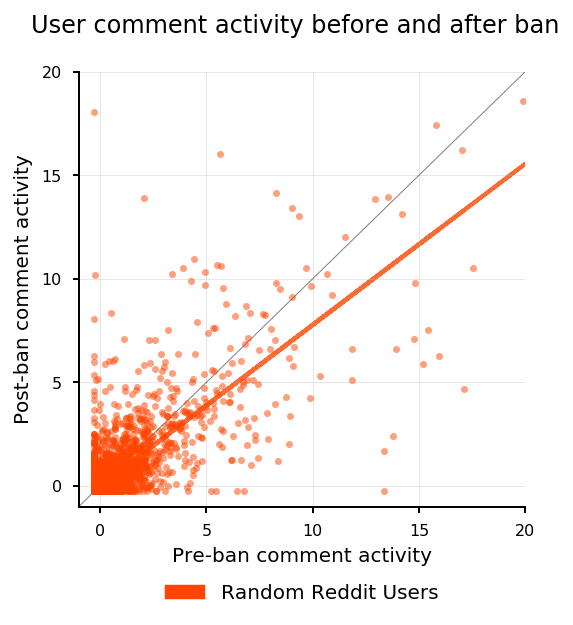} & \includegraphics[width=0.32\textwidth]{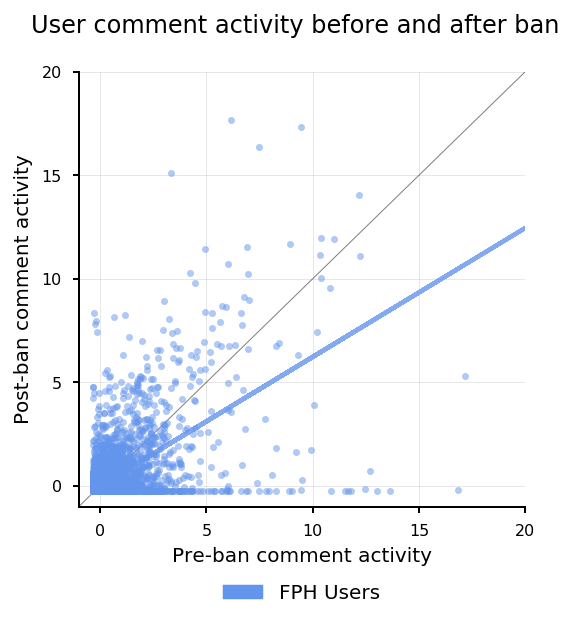} & \includegraphics[width=0.32\textwidth]{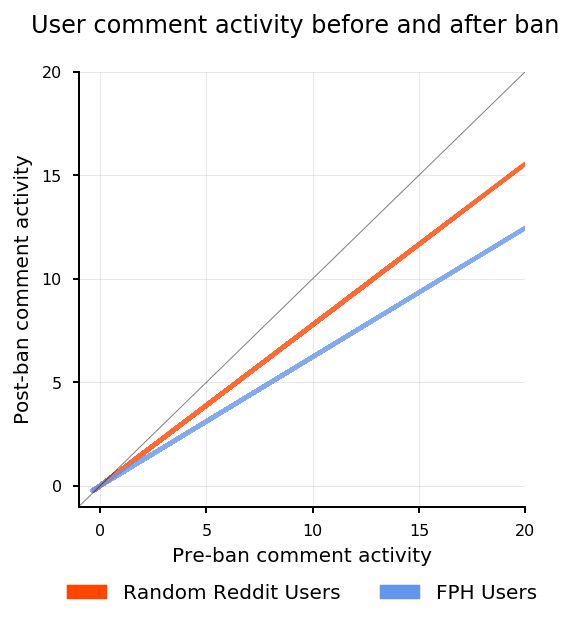} \\
(a) & (b) & (c)
\end{tabular}
\end{center}
\caption{\textbf{The comment activity of FPH and random users before and after the ban.} Scatter plots of pre-ban vs post-ban comment activity of the two user groups with a linear model regression line. A 45 degree regression line indicates equal activity before and after the ban.}
  \label{fig:prevspost}

\end{figure*}

\noindent{\bf The comment activity of the FPH users declined after the ban, when compared to the random users.} Figure \ref{fig:prevspost} presents the effect of the community ban on the two samples individually through scatter plots and their respective regression lines. Figure \ref{fig:prevspost}c presents the regression lines together to emphasize the difference in the proportion of comment activity before and after the ban between the two user groups. From this figure we can infer that the FPH users commented more before the ban, than after, when compared to the random users. This shows that the ban caused a decrease in the Reddit interaction of FPH users. 

\begin{figure}[!ht]
\centering
  \includegraphics[width=0.67\columnwidth]{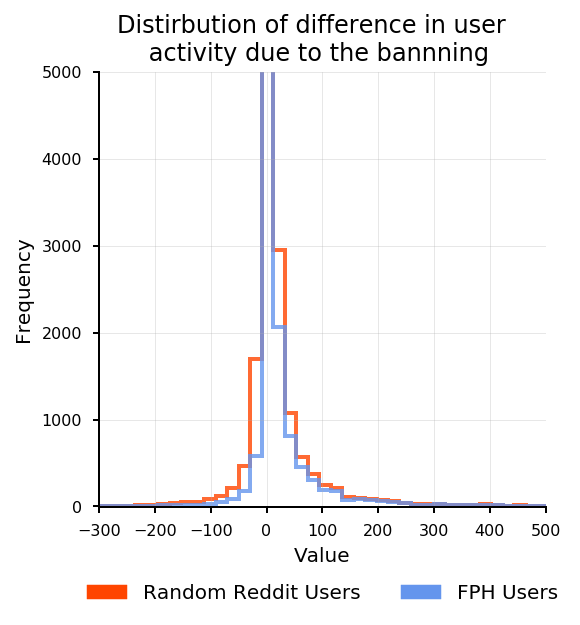}
  \caption{\textbf{Distributions of differences in user comment activity before and after the ban.} Notice the normal-shape of the distribution. The yaxis has been limited to 5,000 for clear depiction of smaller values.}~\label{fig:commdiff}
\end{figure}

To check the significance of this claim, we compare the difference in activity of the two samples. The distribution of the difference in depicted in Figure \ref{fig:commdiff}. Since the distribution is normal, we can perform a paired t-test. For the random sample, the t-test returns a p-value of 0.7, and we fail to reject the null hypothesis that the two samples (pre-ban and post-ban comment activity) are similar. However, for the FPH sample, the paired t-test returns a p-value of 8.7e-197, suggesting that we can reject the null hypothesis and that there is a significant difference in the comment activity before and after the ban for FPH users.

\begin{figure}[!ht]
\centering
  \includegraphics[width=0.67\columnwidth]{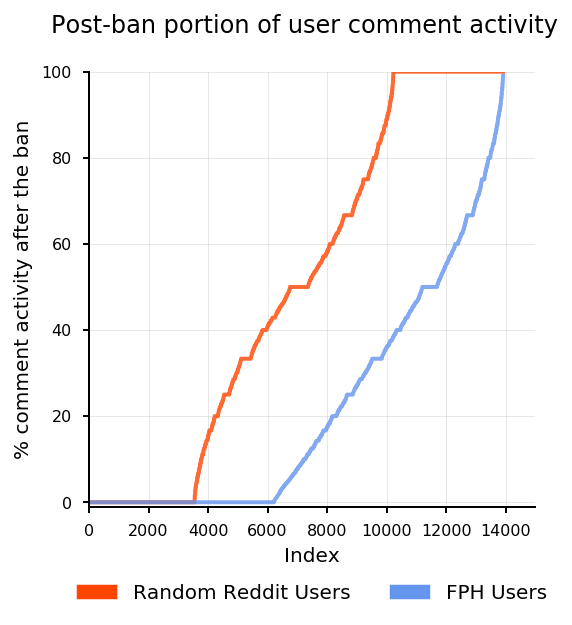}
  \caption{\textbf{Percentage of total comment activity after the ban.} FPH and random users are ranked based on their total comment activity after the ban. At each rank the percentage of post-ban comment activity account out of the total activity is depicted. FPH users are indicated in blue and random users in orange.}~\label{fig:postpercent}
\end{figure}

\noindent{\bf A higher proportion of FPH users completely stopped comment engagement in the post-ban period.} Figure \ref{fig:postpercent} shows the overall trend of Reddit comment activity in the two samples after the ban. We can clearly observe that almost 1.75 times more FPH users became inactive after the banning of the subreddit, when compared to the random sample. Therefore, banning of the subreddit led to higher than average number of accounts with no direct comment engagement.

\begin{figure} [!t]
\centering
  \includegraphics[width=0.67\columnwidth]{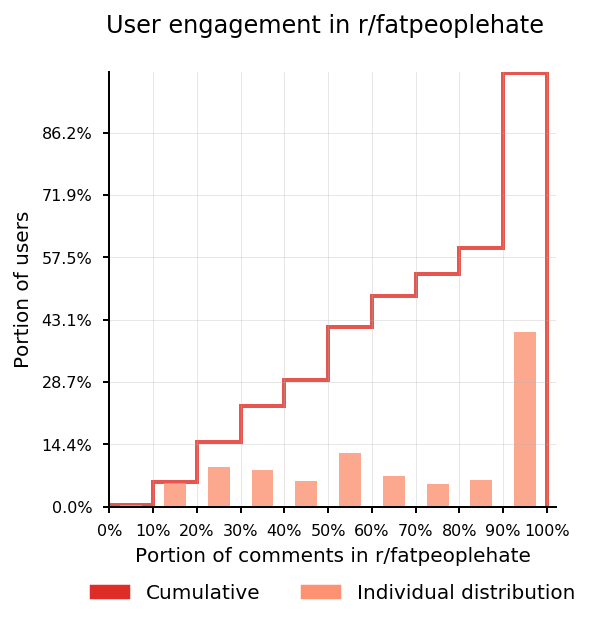}
  \caption{\textbf{Distribution of user engagement in FPH.} Both individual and cumulative distributions are provided.}
  \label{fig:userengagement}
\end{figure}

To delve deeper into this result, we can study the user-engagement of the FPH sample in their subreddit. In Figure \ref{fig:userengagement}, we present the distribution of overall engagement in \texttt{r/fatpeoplehate} by the FPH sample. It is the distribution of users with $x\%$ of their overall comments having been generated in \texttt{r/fatpeoplehate}. From the cumulative distribution, we can assert that more than half the user base made at least half of their comments in \texttt{r/fatpeoplehate}. This shows that our sample of FPH users were highly engaged in \texttt{r/fatpeoplehate} before the ban. Consequently, the ban led to a negative feedback towards this engagement. Hence, we see a steeper decline in the FPH user group, with more FPH users showing no comment engagement than random users.

This effect is also presented in Figure \ref{fig:prevspost}b. Many FPH users have high pre-ban comment activity and almost null post-ban activity. The scattered data points at the bottom of the y-axis and along the length of the x-axis are reflective of the portion of high-activity FPH users who stopped commenting after the ban. Notice a lack of similar pattern of data points in Figure \ref{fig:prevspost}a.

\noindent{\bf A smaller portion of FPH users exhibited negative user behaviour.} Before we present these results, let us preface this by understanding negative user behaviour in context of FatPeopleHate. In this subreddit, users indulged in mocking plus-sized people. While malicious, this behaviour was appreciated and promoted within the subreddit. However, same behaviour would be considered negative outside of the subreddit.

Reddit introduced the banning of the subreddits, on the grounds of user harassment, to control such negative behaviour. While we do not have a measure for gauging hate speech across the entire platform, we can however, study the users that engaged in the harassing subreddit and observe if they continued their actions in other subreddits. 

For this experiment, we use a Reddit feature called `downvote'. Reddit is community-centric and the content generated in the community is monitored by the community as well. Positive user-behaviour is upvoted by the members while negative user behaviour is downvoted. Unfortunately, the downvotes are not reserved solely for harassing behaviour, but can also contain irrelevant or wrong content. Because it is difficult to separate these sub-categories, we inevitably study the overall negative behaviour by FPH users.

\begin{figure} [ht]
\centering
  \includegraphics[width=0.67\columnwidth]{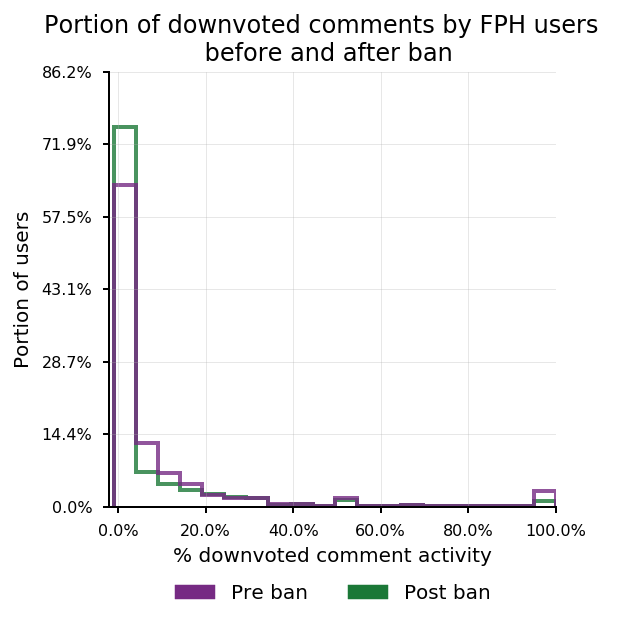}
  \caption{\textbf{Distribution of proportion of downvoted comments in the FPH sample}. It represents that $y\%$ of the users had $x\%$ of their comments downvoted by others.}~\label{fig:badbehaviour}
\end{figure}

Figure \ref{fig:badbehaviour} depicts a histogram of downvoted user activity by FPH users, before and after the ban. It represents that $y\%$ of the users had $x\%$ of their comments downvoted by others. We can make two meaningful observations from this plot:
\begin{enumerate}
    \item For a vast majority of the FPH users, only a small portion of their comments are downvoted.
    \item After the ban, the portion of users with less than 5\% of downvoted comments further increases.
\end{enumerate}

Assuming that had the FPH users exhibited similar behaviour outside of the community, their comments would have been downvoted, the first observation suggests that majority of the FPH users indulged in fat-shaming behaviour, largely within the FPH subreddit. Therefore, we can assert that the \texttt{r/fatpeoplehate} was rather self-contained. The second observation may stem from the fact that a decrease in user comment activity after the ban would also reduce the number of downvoted comments and, therefore, increase the portion of users with low downvoted comment percentage.

Since downvoted comments do not directly suggest harassing / hateful speech, we manually labelled 100 downvoted FPH user comments from before and after the ban. 100 FPH users were randomly selected for this purpose. We found that while 13 of the 100 comments exhibited FatPeopleHate-like behaviour before the ban, the number rose to 25 after the ban. It should be noted that both samples contained comments from subreddits other than \texttt{r/fatpeoplehate} and were manually labelled by an expert.

Therefore, while the overall negative behaviour declined, the hateful behaviour of FPH users was spilling more in other subreddits after the ban.

\noindent{\bf Reddit users explored other subreddits more post-ban, compared to the random sample.} As we have discussed users in the FPH sample were highly engaged in \texttt{r/fatpeoplehate}. This high level of engagement brings with it a high level of association with the subreddit. Banning of the subreddit breaks this association, which can cause users to look for other venues to continue being engaged with the theme of the subreddit. It can also cause users to completely disengage if the subreddit was one of the primary reasons for their use of the platform or continue the engagement with the rest of the platform as before, if they are still interested in rest of the content.

\begin{figure*}
\begin{center}
\begin{tabular}{c c c}
\includegraphics[width=0.32\textwidth]{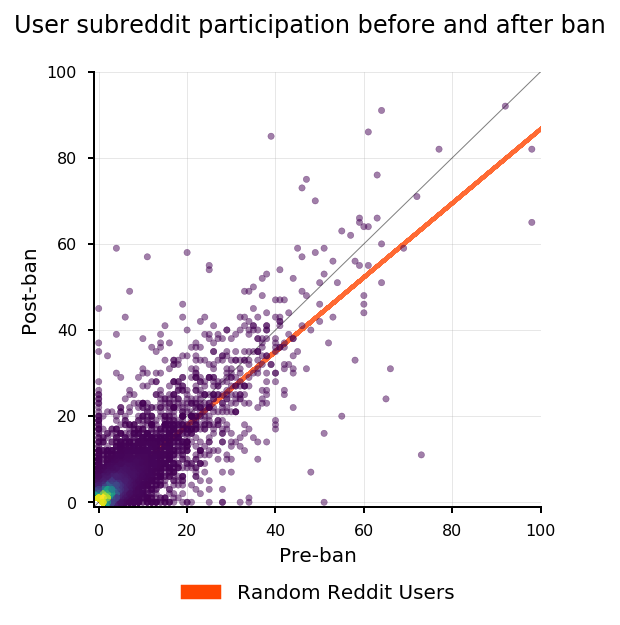} & \includegraphics[width=0.32\textwidth]{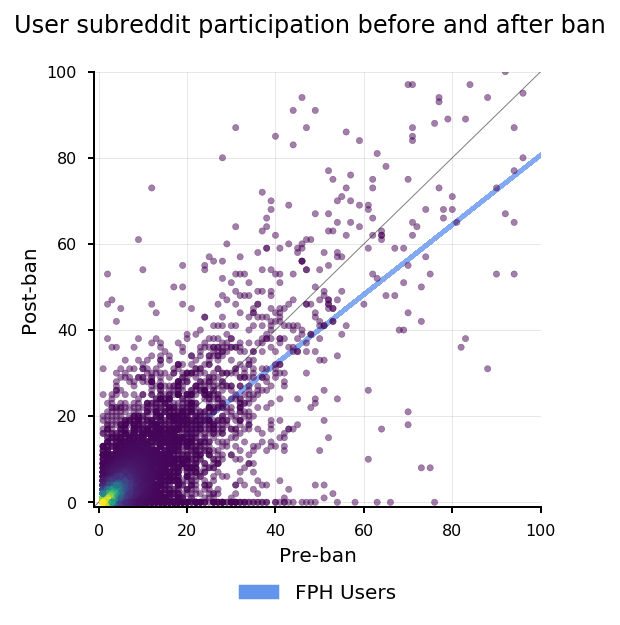} & \includegraphics[width=0.32\textwidth]{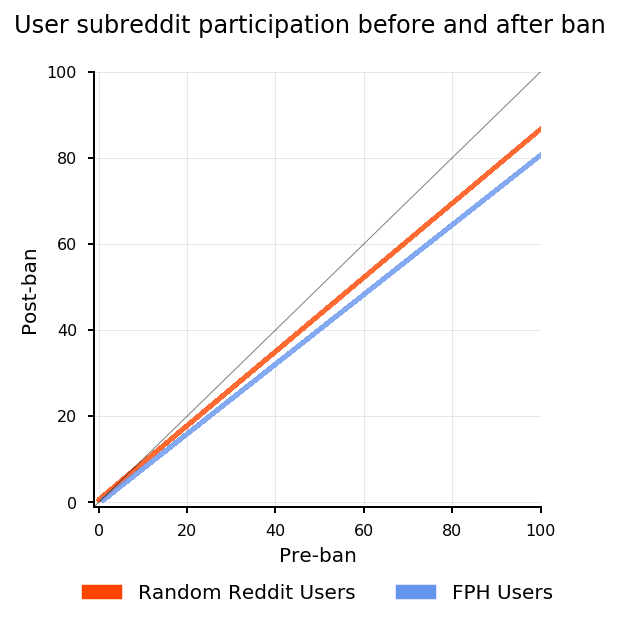} \\
(a) & (b) & (c)
\end{tabular}
\end{center}
\caption{\textbf{User subreddit participation}. The number of subreddits that users commented in before and after the ban. A liner fit is also generated. A 45 degree line indicates equal number of subreddits before and after the ban.}
  \label{fig:prevspossubs}

\end{figure*}

We wanted to study if the ban on their primary subreddit affected multi-subreddit participation of FPH users. Figure \ref{fig:prevspossubs} presents how many subreddits users participated in, before and after banning of \texttt{r/fatpeoplehate}, for both samples. The general trends in the form of  linear regression lines are presented together in Figure \ref{fig:prevspossubs}c. The behaviour for the two samples is quite similar. Therefore, the banning of \texttt{r/fatpeoplehate} did not produce a major difference in how many subreddits FPH users participated in.

\begin{figure}[ht]
\centering
  \includegraphics[width=0.67\columnwidth]{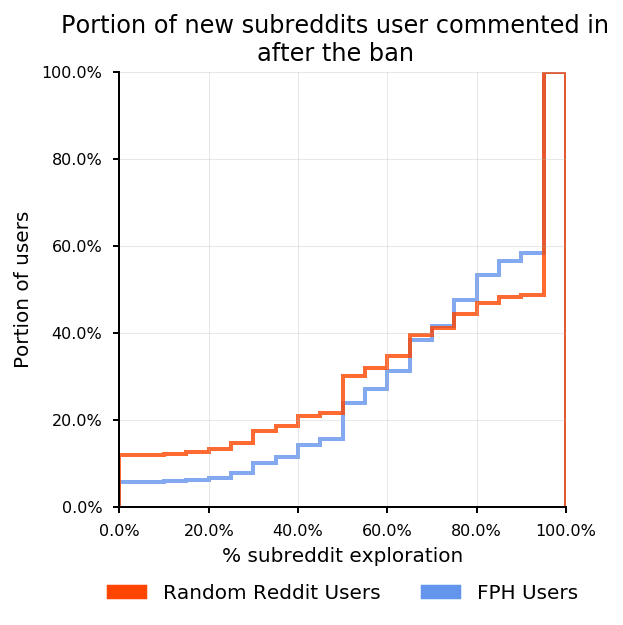}
  \caption{\textbf{Cumulative new subreddit exploration}. The plot shows the percentage of users with atmost x\% of new subreddits after the ban.}
  \label{fig:subexplore}
\end{figure}

While the FPH users continued to participate in an average number of subreddits, we investigated if they were exploring Reddit by participating in subreddits they were not participating in, before the ban, or were they less likely to participate in newer subreddits due to the loss of a subreddit they strongly associated with. In Figure \ref{fig:subexplore}, we observe that 80\% of FPH users were now participating in at least 50\% new subreddits after the ban, as compared to 40\% for the random sample. Further more, the FPH sample had a larger portion of users with high subreddit exploration than the random sample. Therefore, FPH users were participating in more new subreddits after the ban. 

\noindent{\bf FPH sample continued to be interested in the subreddits it was previously active in.} In Table \ref{tab:preposttop20}, we present the subreddits that were popular amongst the users of FPH user sample. From the two lists, we can ascertain that FPH users were still highly interested in the subreddits they used to frequent before. So, even through they were exploring new subreddits, the group as a whole was still actively participating in the subreddits they used to frequent before the ban. However, it is important to adress the fact that these subreddits are popular across Reddit and therefore remain popular after the ban as well.

\begin{table}
\centering
\begin{tabular}{l l}
\small{Pre-ban}                 & \small{Post-ban}             \\
\hline
\small{	fatpeoplehate	}	&	\small{	\textbf{AdviceAnimals}	}	\\
\small{	\textbf{WTF}	}	&	\small{	\textbf{WTF}	}	\\
\small{	\textbf{AdviceAnimals}	}	&	\small{	\textbf{fatlogic}	}	\\
\small{	\textbf{fatlogic}	}	&	\small{	\textbf{TumblrInAction}	}	\\
\small{	TalesofFatHate	}	&	\small{	\textbf{pcmasterrace}	}	\\
\small{	\textbf{TumblrInAction}	}	&	\small{	\textbf{BlackPeopleTwitter}	}	\\
\small{	\textbf{pcmasterrace}	}	&	\small{	\textbf{trashy}	}	\\
\small{	\textbf{BlackPeopleTwitter}	}	&	\small{	\textbf{punchablefaces}	}	\\
\small{	\textbf{trashy}	}	&	\small{	\textbf{trees}	}	\\
\small{	\textbf{trees}	}	&	\small{	technology	}	\\
\small{	\textbf{ImGoingToHellForThis}	}	&	\small{	\textbf{ImGoingToHellForThis}	}	\\
\small{	\textbf{cringepics}	}	&	\small{	\textbf{cringepics}	}	\\
\small{	\textbf{4chan}	}	&	\small{	KotakuInAction	}	\\
\small{	\textbf{fatpeoplestories}	}	&	\small{	\textbf{relationships}	}	\\
\small{	\textbf{punchablefaces}	}	&	\small{	\textbf{fatpeoplestories}	}	\\
\small{	thebutton	}	&	\small{	\textbf{4chan}	}	\\
\small{	FitshionVSFatshion	}	&	\small{	politics	}	\\
\small{	atheism	}	&	\small{	conspiracy	}	\\
\small{	AdiposeAmigos	}	&	\small{	SubredditDrama	}	\\
\small{	\textbf{relationships}	}	&	\small{	Tinder	}	
\end{tabular}
\caption{\textbf{Popular subreddits amongst the FPH users before and after the ban.} The subreddits that are shared by the two lists are in bold font.}~\label{tab:preposttop20}
\end{table}

\noindent{\bf FPH users tried to actively create alternative subreddits for their banned community.} The FatPeopleHate community did not readily accept the banning of their subreddit. They actively tried to circumvent the banning by creating new subreddits to act as an alternative to \texttt{r/fatpeoplehate}. We were able to find 99 such subreddits created in the wake of the ban. Due to limited space we can only provide a few examples: \texttt{r/FatpeoplehateX}, \texttt{r/CandidDietPolice}, \texttt{r/ObesePeopleDislike}, etc. Reddit admins were able to control this surge and banned a majority of these alternatives as well. The ones that were not banned did not get much traffic, most likely due to not being known to the community. Therefore, while FatePeopleHate community reacted by creating a multitude of alternatives, Reddit administration was comprehensive in their policy and banned a majority of them. It is important to note that it is likely that other alternatives were also created and are not present in the list. In case it did happen, the created subreddit is not common knowledge.

\noindent{\bf Effect on weight-related communities}

From the list of subreddits popular with FPH users, we notice \texttt{r/fatlogic} and \texttt{r/fatpeoplestories} to have themes similar to \texttt{r/fatpeoplehate}. While both subreddits make it clear that they are not associated with \texttt{r/fatpeoplehate}, both of them have content that mocks plus-sized individuals. 

For each subreddit, we study the volume of downvoted comments, deleted/removed comments and all comments, made in the respective subreddits during the period of analysis. 

\begin{figure*}
\begin{center}
\begin{tabular}{c c c}
\includegraphics[width=0.31\textwidth]{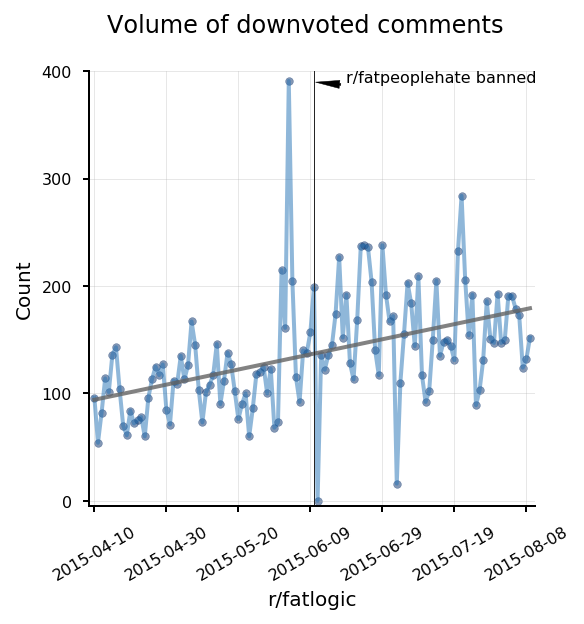} & \includegraphics[width=0.31\textwidth]{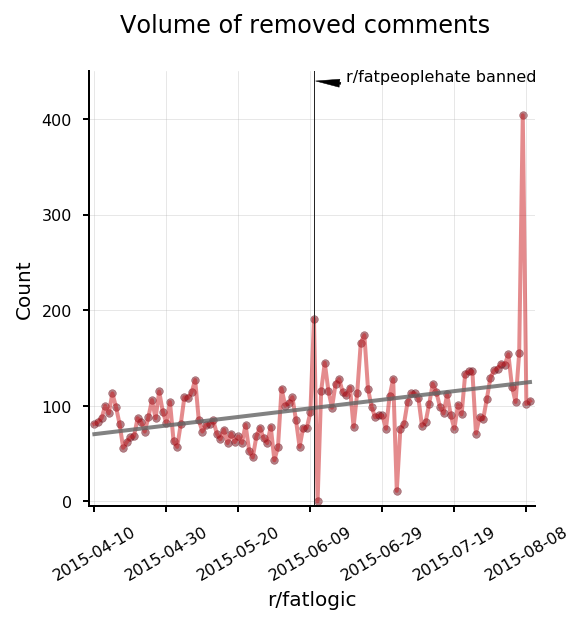} & \includegraphics[width=0.31\textwidth]{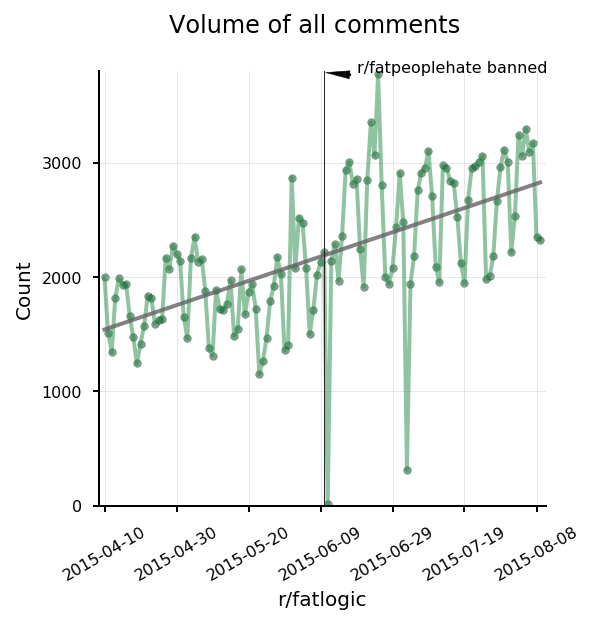} \\
(a) & (b) & (c)
\end{tabular}
\end{center}
\caption{\textbf{Analysis of fatlogic}. There is a dip in volume right after the ban.}~\label{fig:fatlogic}
\end{figure*}

For \texttt{r/fatlogic} (Figure \ref{fig:fatlogic}), we notice zero activity following the ban of \texttt{r/fatpeoplehate}. On further investigation, we found that moderators of the subreddit made it private \cite{maybesaydie2015very}. In a post made by the moderators, they explain how in wake of the banning, the subreddit was flooded with FatPeopleHate. Being overwhelmed by the traffic, and not wanting to be banned if mistakenly perceived as the new home for FatPeopleHate, the mods temporarily made the subreddit inaccessible. An excerpt from the post is provided below:

\blockquote{
``And then, on the morning of June 10, 2015, /r/fatpeoplehate was banned by reddit, We were given no warning and we immediately were overcome with posts from indignant fph subscribers, posts worrying if we were next and it became way too much work for a our small mod team to handle so we went private for a few days. ''
}

Therefore, post-ban, FPH users surged into \texttt{r/fatlogic}, who were able to check this migration by using their moderator privileges to private the subreddit. 

\begin{figure*}
\begin{center}
\begin{tabular}{c c c}
\includegraphics[width=0.31\textwidth]{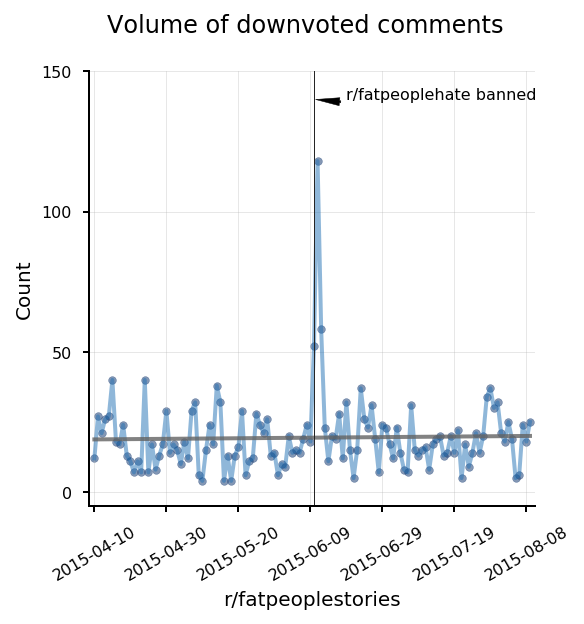} & \includegraphics[width=0.31\textwidth]{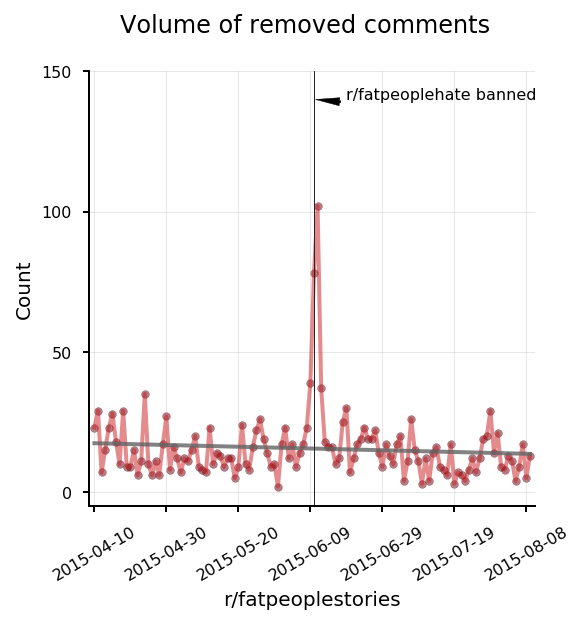} & \includegraphics[width=0.31\textwidth]{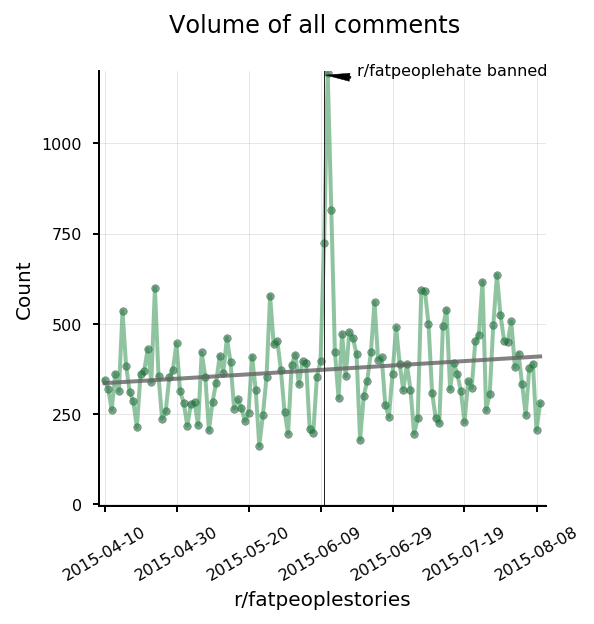} \\
(a) & (b) & (c)
\end{tabular}
\end{center}
\caption{\textbf{Analysis of fatpeoplestories}. The volume parameters resume normal values after the surge.}~\label{fig:fatpeoplestories}
\end{figure*}

\texttt{r/fatpeoplestories} (Figure \ref{fig:fatpeoplestories}), however, kept their subreddit public. Although, we do witness a spike in the total number of comments posted on the subreddit, following the ban, it is short-lived and normal traffic patterns resume soon after. We see a spike in both the number of downvoted comments and the number of deleted comments. We can assert that the community actively checked the surge in traffic by downvoting more than usual. Similarly, moderators also actively kept the migration in check by deleting comments that went against the subreddit.

\begin{figure*}
\begin{center}
\begin{tabular}{c c c}
\includegraphics[width=0.31\textwidth]{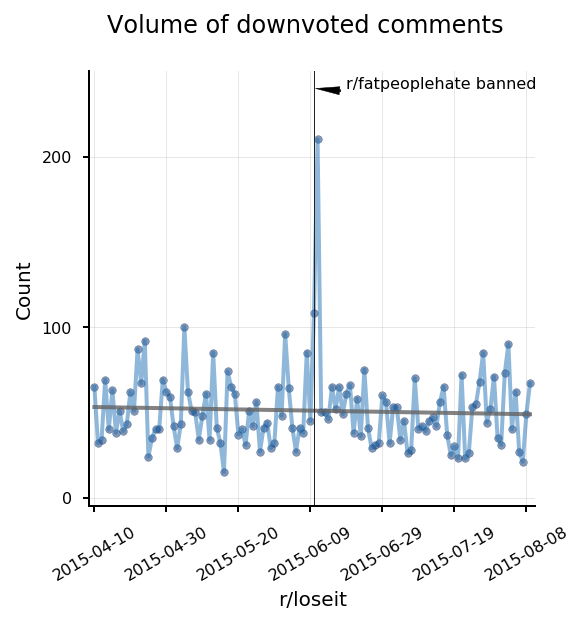} & \includegraphics[width=0.31\textwidth]{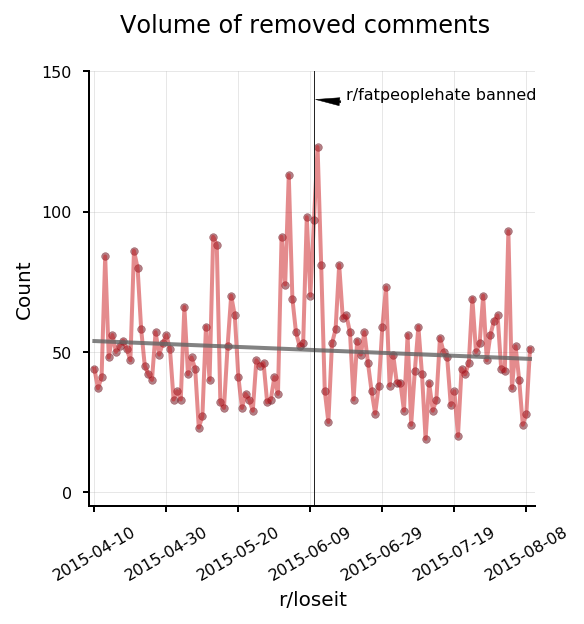} & \includegraphics[width=0.31\textwidth]{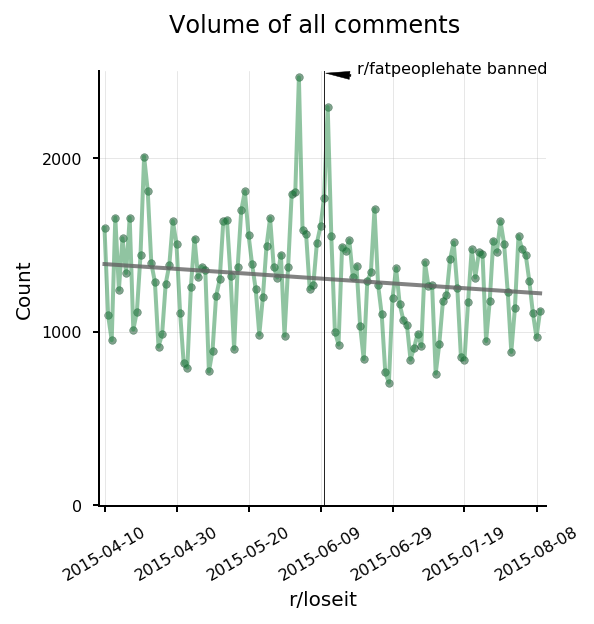} \\
(a) & (b) & (c)
\end{tabular}
\end{center}
\caption{\textbf{Analysis of loseit}. There is a higher than usual count of downvoted and deleted comments right after the ban.}~\label{fig:loseit}
\end{figure*}

Finally, we also study a subreddit, which while being about weight, is on the other end of the spectrum as FPH. \texttt{r/loseit} is a supportive subreddit for users who want to lose weight. In Figure \ref{fig:loseit} we observe that a higher than normal number of comments were posted on the subreddit following the FPH ban. However, since the demographic of the community is plus-sized people, these comments also include the discussion on the banning of a subreddit that was abusive to them. Nonetheless, we also notice that a higher than average number of comments were deleted and that there is a major spike in the volume of downvoted comments. Therefore, we can say that \texttt{r/loseit} did face increased negative user behaviour following the ban of FatPeopleHate. Yet, in this case too, the community and the moderators were actively checking for such behaviour and soon after, normal characteristics resumed.

\section{Discussion}

In this project, we studied the after-effects of banning of the Reddit community r/FatPeopleHate on (1) its active members and (2) other relevant subreddits.
We discovered that the banning had an adverse effect on the members of the FPH community. Not only did we observe a significant decrease in the user comment activity from FPH users after the ban, we also found that a larger portion of the users had completely ceased commenting on Reddit in any way. It is important to note that while Reddit was able to sustain this disengagement, it is not necessarily a positive outcome in every scenario. For new platforms or niche platforms with limited user base, user disengagement, if in large enough numbers, can lead to the demise of the platform.
While we cannot directly state whether the overall FPH-related hate speech decreased after the ban, we can say that the members of the FPH community were less active on Reddit following the ban
The FPH community did try to actively counteract the ban by creating alternative subreddits for their disbanded community. However, Reddit administration was persistent and banned many these subreddits offshoots. Even though some of these alternatives managed to escape the ban, they failed to grow. Therefore, FPH users were unable to recreate their confined space of FPH-replated activities.
Furthermore, we observed that apart from continued interest in major subreddits, in the days following the ban, FPH users started visiting \texttt{r/fatlogic}.  Presumably this happened because, among active subreddits it is most closely related to the theme of FatPeopleHate. However, the moderators of fatlogic were not keen on this surge and made the subreddit private. Other relevant subreddits, namely \texttt{r/fatpeopelstories} and \texttt{r/loseit} also witnessed a spike in former FPH user activity after the ban was enacted. This spike was accompanied with spikes in the number of comments that were downvoted by the community and the number of comments that were deleted, in part, by moderators. In short, FPH users, unable to recreate their community, were also unable to co-opt established communities. Their surge in such communities was mitigated by a combined action by other users and moderators.
For the users that were still engaged, fewer of their post-ban comments were downvoted by other members of the platform. However, a higher portion of the downvoted comments contained FPH-related content. In other words, while fewer comments by the FPH members were getting downvoted, more of these comments contained hateful speech. 
Overall, our findings strongly suggest that banning a community to counter generation of hateful content can succeed in situations where such content is mostly confined within that community and other communities are appropriately empowered with means to curate their content and membership. Of course, the story might be quite different for harassing communities whose negative behavior has already infiltrated other communities. The latter can prove more difficult to curb at the administrative level but might be restrained at the user/moderator level through regular action, in view to the fact that other subreddits demonstrated a remarkable resistance to the incursion of FPH content. This suggests a banning strategy needs to be accompanied by strong moderation of other communities and negative reinforcement of hateful-content by other users for it to be a success.

\section{Conclusion}

In this article, we studied the outcome of a common practice among online forums of banning unsavory activities. While it is more common for group moderators to ban unwanted users, we specifically studied the banning on a larger scale, that of an entire sub-community. Banning of a large and popular sub-group is not a risk-free endeavor. A priori banning hardly guarantees that the content produced by the community will go away. Nevertheless, in this case, we found that Reddit employed a strategy in combination with a suite of user and administrator moderation and curation tools that made the banning effort successful. We observed that banning discouraged users from interacting with the platform, especially when accompanied with appropriate measures are to persist negative reinforcement. Presumably, banning of a community removes an association that users had with the platform, leading to future dis-engagement. We believe that this suggests a concrete approach which could help other platform maintainers build and manage civil and constructive online spaces.

In this case, Reddit was successful in banning FatPeopleHate. However, there are factors that other platform maintainers should be mindful of before enacting similar policies. Notably, the resulting user disengagement can prove to be critical to the platform itself. Furthermore, depending on the scale of the platform, it is going to be harder for the platform maintainers to control the creation of offspring sub-communities. For example, Facebook deals with a much larger user base and banning a popular page would result in the creation of many more alternative pages. It would also be harder to control groups that have off-platform / real-life connections since they can persist. Nevertheless, banning the community can provide significant control to platform maintainers.

\bibliographystyle{aaai}
\bibliography{icwsm}

\end{document}